\UseRawInputEncoding

\documentclass[12pt]{article}
\usepackage[cmtip,arrow]{xy}
\usepackage{pb-diagram,pb-xy}
\usepackage{amsmath}
\usepackage[psamsfonts]{amssymb}
\usepackage{cmmib57}
\usepackage{amsmath,amssymb,amscd}
\usepackage{tikz-cd}
\newcommand{\bPf}{\par\vspace*{-4pt}\indent{\sc Proof.}\enskip}
\newcommand{\ePf}{\medskip}
\def\QED{\hskip0.1em\hfill\null\ \null\nobreak\hfill\kern3pt\vbox{\hrule\hbox
   {\vrule\kern1pt\vbox{\kern1.7pt\hbox{$\scriptscriptstyle{QED}$}
    \kern0.2pt}\kern1pt\vrule}\hrule}}

\def\END{\hskip0.1em\hfill\null\ \null\nobreak\hfill\kern3pt\vbox{\hrule\hbox
   {\vrule\kern1pt\vbox{\kern1.7pt\hbox{$\,\,\,\vspace{5pt}$}
    \kern0.2pt}\kern1pt\vrule}\hrule}}
\newtheorem{theorem}{Theorem}[section]
\newtheorem{lemma}[theorem]{Lemma}
\newtheorem{corollary}[theorem]{Corollary}
\newtheorem{proposition}[theorem]{Proposition}
\newtheorem{remark}[theorem]{Remark}
\newtheorem{definition}[theorem]{Definition}
\newtheorem{example}[theorem]{Example}

\DeclareMathOperator{\byd}{{\raisebox{.1ex}{:}{=}}}
\newcommand{\bCd}{\beq \begin{CD}}
\newcommand{\eCd}{\end{CD}\eEq}
\newcommand{\bcd}{\beq \begin{CD}}
\newcommand{\ecd}{\end{CD}\eeq}
\newcommand{\ben}{\begin{enumerate}}
\newcommand{\een}{\end{enumerate}}
\newcommand{\bEq}{\begin{eqnarray}}
\newcommand{\eEq}{\end{eqnarray}}
\newcommand{\beq}{\begin{eqnarray*}}
\newcommand{\eeq}{\end{eqnarray*}}
\newcommand{\bDf}{\begin{definition}\em}
\newcommand{\eDf}{\end{definition}}
\newcommand{\bLm}{\begin{lemma}}
\newcommand{\eLm}{\end{lemma}}
\newcommand{\bPr}{\begin{proposition}}
\newcommand{\ePr}{\end{proposition}}
\newcommand{\bTh}{\begin{theorem}}
\newcommand{\eTh}{\end{theorem}}
\newcommand{\bCr}{\begin{corollary}}
\newcommand{\eCr}{\end{corollary}}
\newcommand{\bRm}{\begin{remark}\em}
\newcommand{\eRm}{\end{remark}}
\newcommand{\bEx}{\begin{example}\em}
\newcommand{\eEx}{\end{example}}

\newcommand{\ie}{{\em i.e$.$} }
\newcommand{\eg}{{\em e.g$.$} }
\newcommand{\R}{I\!\!R}

\newcommand{\der}{\partial}

\newcommand{\cE}{\mathcal{E}}

\newcommand{\cI}{\mathcal{I}}

\newcommand{\cL}{\mathcal{L}}

\newcommand{\cR}{\mathcal{R}}

\newcommand{\cW}{\mathcal{W}}

\newcommand{\wed}{\wedge}

\newcommand{\lam}{\lambda}
\newcommand{\sig}{\sigma}

\newcommand{\ome}{\omega}

\title{\large{ {\bf 
Lepage equivalents for second order Lagrangians and applications: $2D$ modified higher order Boussinesq-type equations
}}
}
\author{
{\normalsize Marcella Palese\footnote{Corresponding Author}}
\\ {\footnotesize Department of Mathematics,
University of Torino}
\\
{\footnotesize via C. Alberto 10, 10123 Torino, Italy, e--mail: 
{\sc marcella.palese@unito.it}} 
\\
{\normalsize Fabrizio Zanello}
\\
{\footnotesize 
Institute of Mathematics,
University of Potsdam }\\
{\footnotesize Campus Golm, Haus 9
Karl-Liebknecht-Str. 24-25, Potsdam OT Golm,
Germany, }\\ {\footnotesize e--mail: {\sc fabrizio.zanello@uni-potsdam.de}}
}
\date{}
\begin{document}
\maketitle

\begin{abstract}  
In the frame of the Lagrangian formalism on $r$-order prolongations of fibered manifolds and related structures such as (prolongation of) projectable vector fields, (sheaves of) differential forms and contact structures, we propose a Lagrangian two-field derivation of $2D$ modified Boussinesq equations, obtained as coupled systems of Euler--Lagrange (E-L) equations for the two fields. By means of a recursive formula involving geometric integration by parts formulae, we construct extended `full' equivalents of such Lagrangians, in particular of Krupka--Betounes type, by which the equations are obtained straightly as the $1$-contact component of their exterior differential.  As a main result we find {\em new $2D$ fourth- and sixth-order modified Boussinesq-type equations}, containing mixed terms in both the spatial variables $x$ and $y$.
As a byproduct, we also obtain a {\em $2$-field variational characterization} of the stationary reduction of the moving-frame (according to Bogdanov and Zakharov) KP equation. 

\end{abstract}




\section{Introduction}

As pointed out by Boussinesq himself \cite{Boussinesq1872}, in the first approximation, his model
reduces exactly to the linear wave equation. Thus it is not surprising that, in its different variants, modifications and extensions, the Boussinesq equation  
describes,  indeed, a  wide range of real world phenomena which mimic, with suitable variants, the {\em motion of long dispersive shallow water waves}, \ie physical phenomena in diverse fields of sciences such as biology, condensed matter physics, plasma physics, plasma waves, fluid mechanics,  oceanography, cosmology and fundamental forces of nature.
This explains why various Lagrangian formulations and analysis of various generalizations or modifications and extensions to $(2+1)$ dimensions ($2D$) of the classical $(1+1)$ Boussinesq equation have been proposed and investigated within many different approaches. 

In particular, a $(1+1)$ modified Boussinesq equation whereby the fourth-order term is a mixed term containing two time derivatives and two space derivatives has been considered;
see \eg \cite{RGRB19} and references therein, and \cite{ChWaWa04,WaCh06}. 
A further interesting modification of such a model is represented by a $(1+1)$ {\em sixth-order} Boussinesq-type equation modelling long gravity-capillary surface waves with a
short amplitude, propagating in both directions in shallow water; see \eg \cite{RGB16}.
For an extended and generalized Boussinesq equation is usually intended a $(2+1)$ (or $2D$) extension of the $(1+1)$-dimensional model obtained by adding a second order term in a further spatial variable whereby the nonlinear term is generalized as the second order spatial derivative of a generic positive power of the field or the nonlinear term is given by a generic function of the first order $x$-derivative of the field;
 see \eg \cite{Lu_etal,MWT89,RZZ14,RGRB19}, and references therein.
In this paper we first frame such models within the Lagrangian formalism on $r$-order prolongations of fibered manifolds and related structures such as (prolongation of) projectable vector fields, (sheaves of) differential forms and contact structures, then we propose an alternative Lagrangian two-field formulation, in such a way that $2D$ modified Boussinesq equations can be obtained as coupled systems of Euler--Lagrange (E-L) equations for the two fields. As a main result we find {\em new $(2+1)$ fourth- and sixth-order modified Boussinesq-type equations}, containing mixed terms in both the spatial variables $x$ and $y$.

The interest of such a second order two-field Lagrangian formulation is the possibility of applying a machinery which associates to the Lagrangian its (in a sense which will be specified later on) `full' equivalent \cite{PaRoZa22,PaZa23}.
Indeed, according to the seminal works of Cartan \cite{Cart22} and Lepage \cite{Lep36}, the E-L  equations can be described by the concepts of sheaves of differential forms and their exterior differential modulo sheaves of contact structures.
Within this perspective, the finite order variational sequence, as a quotient sequence of the de Rham sequence, was introduced and developed by Krupka; see \eg  \cite{Kru15}. 
The problem of the representation of the finite order variational sequence and, specifically, of the arrow representing the Euler--Lagrange map,  has been discussed in terms of the so called {\em interior Euler operator} and corresponding geometric integration by parts; see \eg  \cite{KrMu05} and \cite{PaRoWiMu16,PaRoZa22}.
Indeed, since Lepage equivalents are concerned with the boundary term emerging in the integration by parts of the action integral,
we approached this aspect from a geometric sheaf-theoretical point of view and obtained (local) expressions of Lepage equivalents by means of suitable {\em geometric residual operators}.
In particular a `full' Lepage equivalent, a {\em local} extension to the second order case of the so-called Krupka--Betounes Lepage equivalent 
\cite{Kru77,Betounes84}, was obtained within this formalism \cite{PaRoZa22}. Recently the problem of the globalization of local Lepage equivalents has been tackled in \cite{Sau24}.

In the middle of the 80s, Betounes studied the symmetry theory of such a kind of Lagrangian equivalents \cite{Betounes98}.
He stressed that, while symmetries of Lagrangians are also symmetries of their Lepage equivalents, in general there exist symmetries of the Lepage equivalents which are not symmetries of the Lagrangians; the corresponding conservation laws, and related conserved quantities, are then different. 
This relays on the fact that, depending on the symmetries, the Lagrangian is invariant up to divergences (trivial Lagrangians) which are, in general, of different {\em `nullity'} 
(according to Betounes'  definition, a first order trivial Lagrangian function $L$ has nullity $k$ if $\der^{k+1} L / \der y^{a_{1}}_{i_{1}}\cdots \der y^{a_{k+1}}_{i_{k+1}}= 0$ on each chart, see later for the notation).
In particular, Betounes characterized symmetries transformations of solutions of the E-L equations  (so-called point symmetries) which are also symmetries of the Krupka-Betounes equivalent.
This problem is related to the question as to determine the condition for a Noether--Bessel-Hagen current \cite{Noe18,BeHa21}, associated with a generalized symmetry, to be variationally equivalent to a Noether current for a suitable invariant Lagrangian
\cite{CaPaWi16,PaWi17}, and with the characterization of symmetry transformations of extremals and related higher conserved quantities \cite{AcPa21}.
Invariance of Lepage equivalents provides `improved' Noether conserved quantities, which somehow reveal the invariance of a system as a variational one (in its Lagrangian-equivalent form) rather than barely the invariance of (set of solutions of) equations.

In this work we discuss some aspects concerning 
 `full' extended equivalents   
of Krupka--Betounes type, which {\em differ} from the Poincar\'e-Cartan form by contact terms of higher degree; their exterior differential modulo a suitable contact structure produce the {\em same Euler--Lagrange equations}, they inherit the {\em same symmetries} of the Lagrangian (which turn out to be also symmetry transformations of the dynamics)
 however, in principle, they can define new conserved quantities. 
 Accordingly, the concept of variational symmetries can be extended and generalized as point symmetries (symmetries of the set of solutions of the corresponding Euler--Lagrange equations) having a `full' variational meaning: see \eg \cite{LeMuCz09} for examples of applications in Mechanics, and \cite{Perez19}. 

Besides their variational origin, symmetries and conservation laws provide insights on integrability, existence of soliton solutions, B\"acklund transformations, Wahlquist--Estabrook prolongation algebraic structures associated with nonlinear systems. Notably, the geometric formulation of conservation laws in terms of a generalization of the concept of a {\em connection on the bundle of fields} turns out to be relevant also in real world applications, see \eg \cite{Pa16_Eco}.

Concerning the physical systems under investigation, a $(1+1)$ Boussinesq type equation was introduced in \cite{Zak74}, as an integrable system describing a nonlinear string interpreted as the continual analog of the Fermi-Pasta-Ulam problem by a chain with a quadratic nonlinearity. 
Applications to Conformal Field Theory (CFT) in the context of quantum Boussinesq theory can be also found in the literature, see \eg 
\cite{BaHiKh02,BeCh93,Mat88}. 
Furthermore, encompassing, among other, vortex-like phenomena,  the $(2+1)$-dimensional models appear to be relevant.

It is well known that the integrability theory for high dimensional nonlinear systems is in many aspects nontrivial; see \eg \cite{Pa16_Eco,PaWi02,PaWi11,PaWi20} and references therein. 
Notably, also the extension of Lagrangians in the Lepage sense  
turns out to involve nontrivial structures, especially concerning uniqueness or globality, see the discussion in \cite{Sau24}.  
In the context of integrable nonlinear systems, Noether conserved currents have been related with momenta, (multi)symplectic and Hamiltonian structure(s), as well as Dirac structures (see, \eg \cite{Mag78}). Within this perspective the Lepage equivalent approach enables to include the {\em`full'  variational content} of integrability properties based on the study of  symmetries and conservation laws. The aim of this paper is to explore the interplaying of the two approaches.

 \section{Framework: geometry and symmetry}

Fibered manifolds and their jet prolongations represent a convenient mathematical framework for mechanics and field theories, describing Lagrangian systems of different orders, and depending on many independent variables within a sufficiently general and unified geometric framework. 

In what follows, we shall consider a smooth fibered manifold $\pi: Y \to X$, with $\dim X = n$ and $\dim Y = n+m$.
 If  $(x^i, y^{\alpha} )$, are local adapted coordinates, the physical  fields are described by sections $(x^i, \gamma^{\alpha} (x^{j}) )$.
 
Let $J^rY$ (the $r$-jet prolongations $\pi_r : J^r Y \to X$, with $r\ge 0$ any integer) be the manifold of points $j^r_x\gamma$, \ie of  equivalence classes
of $C^r$ local sections $\gamma$ of $\pi$ with source $x\in X$ and target $y=\gamma(x)\in Y$  with the same value at $x$ and the same partial derivatives at $x$ up to the order $r$. 
Induced local adapted coordinates are $(x^i, y^{\alpha}, y^{\alpha}_{j_1, \ldots, j_r})$; partial derivatives of fields are then described by holonomic sections $(x^i, \gamma^\alpha (x^i), \der \gamma^\alpha (x^j )/ \der x^{j_1}, \ldots )$, see \eg \cite{Sau89} for more detail. 
For $n =1$, the manifold $Y$ is a space of events for mechanical systems of $m$ degrees of freedom, and local sections of $\pi$ are graphs of curves, so usually $X = \R$; for $n  > 1$, the local sections of $\pi$ describe physical fields over the manifold $X$.

Due to the affine bundle structure of $\pi^{r+1}_{r} : J^{r+1} Y \to J^{r} Y$, we have a natural splitting $ J^rY\times_{J^{r-1}Y}T^*J^{r-1}Y = J^rY\times_{J^{r-1}Y}(T^*X \oplus V^*J^{r-1}Y)$ which induces natural splittings into horizontal and vertical parts of projectable vector fields, forms, and of the exterior differential on $J^rY$.

A differential $q$-form $\omega$ on $J^r Y$ is called {\em contact} if $(j^r\gamma)^*\omega = 0$ for every local section $\gamma$ of $\pi$. 
Contact forms on $J^r Y$ constitute an ideal in the exterior algebra, called the {\em contact ideal} of order $r$, which is generated by contact $1$-forms and their exterior derivatives.
A contact form $\omega$ is called $k$-contact if for every vertical vector field $\Phi$ the contraction $\Phi\rfloor \omega$ is $(k-1)$-contact. 
A differential $q$-form $\omega$ on $J^r Y$ is called {\em horizontal}, or {\em $0$-contact}, if $\Phi\rfloor \omega = 0$ for every vector field $\Phi$ on $J^r Y$, vertical with respect to the projection onto $X$. 
For $q > n$, every $q$-form on $J^r Y$, when lifted to $J^{r+1} Y$, is contact and, for $q\leq n$, is a sum of a unique horizontal and contact form.
The {\em total (formal) derivative} is locally expressed as 
\beq
d_i = \frac{\partial}{\partial x^i} 
|_{  j^r_x\gamma}+\sum_{k=0}^{r}\sum_{j_1\le j_2\le\dots\le j_k}y^\sigma_{j_1j_2\dots j_ki}\frac{\partial}{\partial y^\sigma_{j_1j_2\dots j_k}} 
|_{  j^r_x\gamma} \,.
\eeq
The {\em contact $1$-forms} (or Pfaff forms)
\beq
 \omega^\sigma_{j_1j_2\dots j_k}  = dy^\sigma_{j_1j_2\dots j_k} - y^\sigma_{j_1j_2\dots j_k i}dx^i  \, ,
\eeq
encode partial derivatives, roughly speaking they are a sort of {\em field's `virtual displacements'} analogous to Mechanics' ones.

To fix the notation, we shall denote the volume and lower degree density forms as
$ds=dx^1\wedge\ldots\wedge dx^n$,
$\quad ds_{ij}=\frac{\partial}{\partial_{x^j} } \rfloor \frac{\partial}{\partial_{x^i} }\rfloor ds \,, \quad $ and so on.

The affine bundle structure of $\pi^{r+1}_r \colon J^{r+1}Y\longrightarrow J^rY$ induces natural morphisms $p_i \colon\Omega^r_q W \longrightarrow\Omega^{r+1}_q W$, $0\leq i \leq q$ of exterior algebras  \cite{Kru73,Kru15}.
For any $\rho\in\Omega^r_qW$,  $q\ge 0$, we have the unique decomposition
$(\pi^{r+1}_r)^\ast\rho=  \sum_{i=0}^q p_i\rho $
into a sum of $i$-contact forms, $0 \leq i \leq q$; $h = p_0$  denotes the `horizontalization'. 
The form $p_i \rho$ is called the {\em $i$-contact component of $\rho$}. 
Note that if $q \geq n+1$ then $\rho$ is contact, and it is {\em at least $(q-n)$-contact}, \ie the contact components $p_0, p_1, \dots, p_{q-n-1}$ of $\rho$ are equal to zero. If $p_{q-n}\rho = 0$, $\rho$ is a {\em strongly contact} form.

The corresponding splitting of the pull-back of the exterior differential reads
\beq
(\pi^{r+2}_r)^\ast d\rho=d_H\rho+d_V\rho 
: = \sum_{i=0}^q p_i dp_i\rho + \sum_{i=0}^q p_{i+1}dp_i\rho \,,
\eeq
where the {\em horizontal differential} is $d_H = hd$, and the {\em vertical differential} $d_V$ is the contact part of  
$(\pi^{r+1}_{r})^* d$.

The affine bundle structure of $\pi^{r+1}_{r}$ induces a splitting of vector fields along the projection $\pi^{r+1}_{r}$ in a horizontal and a vertical part.  \\
Let
$
\xi = \xi^i \frac{\partial}{\partial x^i} + \sum_{|J|=0}^r \xi^\sigma_{J} \frac{\partial}{\partial y^\sigma_J}
$
be a vector field on $J^rY$, where $J$ is a multi-index, $0 \leq |J| \leq r$.  The splitting $\xi \circ \pi^{r+1}_{r} = \xi_H + \xi_V$,  is then given in local coordinates as 
\beq
\xi_H = \xi^i d_i \,, \quad 
\xi_V =  \xi\circ \pi^{r+1}_{r} - \xi_H = \sum_{|J|=0}^r (\xi^\sigma_{J}  - y^\sigma_{Ji}\xi^i) \frac{\partial}{\partial y^\sigma_J}\,.
\eeq
Note that both $\xi_H$ and $\xi_V$ are vector fields along the projection $\pi^{r+1}_{r}$ rather than `ordinary' vector fields on $J^rY$; furthermore, if $\xi$ is a contact symmetry, 
$\xi_V$ is a type of characteristic vector field as defined in the theory of Lie symmetry of PDE's \cite{Olver}.
If $\Xi$ is a projectable vector field on $Y$, we write $J^r \Xi_V = (J^r \Xi)_V$ (this can be viewed as a definition of prolongation of $\Xi_V$ which is a vector field along the projection $\pi^{1}_{0}$).   

\smallskip

In the calculus of variations and the theory of differential equations on fibered manifolds, 
the fundamental role is played by the sheaf $\Lambda^r_{n,X}$ of horizontal ($0$-contact) $n$-forms on $J^r Y$, the elements of which are called  {\em Lagrangians} of order $r$, and the sheaf $\Lambda^r_{n+1,1,Y}$ of $1$-contact $(n+1)$-forms on $J^r Y$, horizontal with respect to the projection onto $Y$, the elements of which are called {\em dynamical forms} of order $r$. 

By means of the variation of the action defined by a Lagrangian $\lambda$ 
one obtains a distinguished dynamical form $E_\lambda$, called the {\em Euler--Lagrange form} of $\lambda$ \cite{Kru15}; the components of $E_\lambda$ in every fibered chart are the Euler--Lagrange expressions of the Lagrangian $\lambda$. Remarkably, if $\lambda$ is of order $r$ then $E_\lambda$ is of order $\leq 2r$.
The mapping assigning to every Lagrangian its Euler--Lagrange form is called the {\em Euler--Lagrange mapping}.

Let $k \geq 1$. An $\omega^\sigma$-generated $k$-contact $(n+k)$-form is called a source form of degree $n+k$    .
For $k=1$, dynamical forms are $\omega^\sigma$-generated $1$-contact $(n+1)$-forms, hence, indeed, source forms of degree $n+1$. 
Source forms of degree $n+2$ are called {\em Helmholtz-like forms}. 

A vector field $\Psi$ on $J^rY$ is called a {\em contact symmetry} if it is a symmetry of the contact ideal ${\cal C}_r$, \ie for every contact form $\omega$ the Lie derivative $L_\Psi  \omega$ is a contact form.

For any projectable vector field $\Xi$ on $Y$ the $r$-jet prolongation $\Psi = J^r\Xi$ is a contact symmetry, and conversely, if a contact symmetry $\Psi$ on $J^rY$ is projectable onto $X$ then $\Psi = J^r\Xi$ for some $\Xi$ on $Y$.
The Lie derivative of a $k$-contact form ($k \geq 0$) by a contact symmetry is at least a form of $k$-contact. 

In the following we shall restrict to {\bf{\em $\pi_s$-projectable}   
 contact symmetries}, in particular  $\Psi = J^s\Xi$ for a projectable vector field $\Xi$ on $Y$. 
It is noteworthy that, in this case, the Lie derivative preserves each contact components, \ie for $ k \geq 0$
 \bEq  \label{contLi}
L_{J^{r+1}\Xi} p_k\rho = p_k L_{J^{r}\Xi} \rho \,. 
\eEq
This generalizes the horizontal case as
\beq
L_{J^{r+1}\Xi} (\pi^{r+1}_{r})^* \rho =  L_{J^{r+1}\Xi} 
( \sum_{k=0}^q  p_k  \rho  
) = \sum_{k=0}^q  p_k (L_{J^{r}\Xi} \rho)   \,.
\eeq

\section{Extension of the Poincar\'e--Cartan form to a `full'  equivalent} \label{Section 3}

Let $\rho$ be a $q=n+k$ form,  $k\geq 1$,  there exists a (global) decomposition
\beq
p_k\rho = \mathcal{I}(\rho) + p_kdp_k \mathcal{R}(\rho)
\eeq
where $\mathcal{I}$ is the \textit{interior Euler operator},  $\mathcal{R}$  is the \textit{residual  operator} \cite{KrMu05}.
Note that here  {\em $\mathcal{R}$ contains the} {\bf momenta} {\em associated with the Lagrangian}, and 
the explicit expressions for $\mathcal{I}$ are given by
\beq
\mathcal{I}(\rho)=
\frac{1}{k}\omega^\sigma\wedge\sum_{|J|=0}^r(-1)^{|J|}\text{d}_J(\frac{\partial}{\partial y^\sigma_J}\rfloor p_k\rho)  \,.
\eeq

Let  $p_0\rho \equiv h\rho=\lambda=\cL ds$ so that $d\rho$ is a $q=n+1$ form, \ie {\em $k=1$}.
A representation of the class $[d\rho]$ (modulo contact structure) provides the well known {\em Euler-Lagrange} form (\ie the {\em variational derivatives}, see later for explicit case expressions)
 \beq  
& E_n(h\rho) = {\cal{I}}(d\rho)={\cal{I}}(dh\rho)= \\
& \sum_{|J|=0}^r
(-1)^{|J|} d_J  (\frac{\partial{\cL } }{\partial
y^\sigma_J} )\omega^\sigma \wedge ds = \epsilon_\sigma(\cL )\omega^\sigma \wedge ds. 
\eeq

\subsection{Geometric integration by parts for $k$-contact forms of lower horizontal degree}
In \cite{PaRoZa22} the extension to a similar geometric integration by parts formula has been obtained in  what we call `the lower degree case'.

Indeed, let $W\subseteq Y$ be a an open set and let $\mu\in\Omega^r_{n-s+k}W$ be a $(n-s)$-horizontal $k$-contact $(n-s+k)$-form.
In a local fibered chart $\psi^r=(x^i,y^\sigma,y^\sigma_I)$, we can write:
\beq 
p_k\mu=\sum_{|J|=0}^r\omega^\sigma_J\wedge\eta^J_\sigma\,\in\Omega^{r+1}_{n-s+k}W
\eeq
where $\eta^J_\sigma$ are local $(n-s)$-horizontal $(k-1)$-contact $(n-s+k-1)$-forms defined on $W^{r+1}Y 
\byd (\pi^{r+1}_{0})^{-1} W$.

There exists a {\em local}  decomposition
\beq
p_k\mu = \mathfrak{I}(\mu) + p_kdp_k \mathfrak{R}(\mu)
\eeq
where $\mathfrak{I}$  is the {\em lower interior Euler operator},  $\mathfrak{R}$  is  the {\em lower} residual  operator,
$\mathfrak{I}(\mu)$ is a $k$-contact source form and  $\mathfrak{R}(\mu)$ is a local $(n-s-1)$-horizontal $k$-contact $(n-s-1+k)$-form (see \cite{PaRoZa22} for further detail).

\bRm
It is of interest for the applications given in this paper the following local characterization
for the case  {\em $k=1$} \cite{PaRoZa22}.
Let $p_1\rho=\sum_{|L|=0}^r\omega^\sigma_L\wedge\eta^L_\sigma\,\in\Omega^{r+1}_{n-s+1}W$, with 
\beq
\eta^L_\sigma=A^{i_1\dots i_sL}_\sigma \wedge ds_{i_1\dots i_s}
\eeq
where $A^{i_1\dots i_s L}_\sigma$ are functions defined on $W^{r+1}$.
We have  the decomposition
\beq
 p_1\rho= \mathfrak{I}(\rho) + p_1dp_1\mathfrak{R} (\rho)
\eeq
 where
\beq
p_1dp_1\mathfrak{R} (\rho)=
d_H [\frac{1}{(s+1)}\sum_{|L|=0}^{r-1} - \hat{t}^{i_1\dots i_siL}_\sigma\omega^\sigma_L\wedge ds_{i_1\dots i_si} ] \,, 
\eeq 
with $\hat{t}^{i_1\dots i_siL}_\sigma$ defined by recursive formulae in terms of $A^{i_1\dots i_sL}_\sigma $  \cite{PaRoZa22}.
\eRm

\subsection{Lepage equivalents}

Lepage $n$-forms introduced by Krupka in 1973 \cite{Kru73} provide a geometric formulation of the intrinsic first variation formula, Noether Theorems and conservation laws for Lagrangian and Hamiltonian theories;  see 
 \eg  \cite{Kru15}.

A $n$-form $\rho$ on $J^rY$ is called a {\em Lepage $n$-form} of
order $r$ if, for every $\pi^{r}_{0}$-vertical vector field $\psi$ on
$J^rY$, $
h(\psi \rfloor d\rho)=0$.
Its horizontal component $h\rho$ is, of course, a Lagrangian on $J^{r+1}Y$.
Every $n$-form {\em defined on $Y$} is a Lepage $n$-form and it defines a Lagrangian on $J^1Y$, which is a polynomial of degree $n$ in the first derivatives.
Furthermore, every {\em closed} $n$-form on $J^rY$ is a Lepage $n$-form which defines a Lagrangian having the zero Euler--Lagrange form.

Note that $\rho$  is a Lepage $n$-form of order $r$ if and only if
$p_1 d\rho$ is a dynamical form, \ie, $p_1d\rho = \cI(d\rho)$; or equivalently if and only if
 ${\pi^{r+1}_{r}}^* d\rho=  E+F$, where $E$ is a dynamical form, and $F$ is at least $2$-contact.
 
Let $\rho \in \Omega^r_n$. By the definition and properties of $\cI$ and $\cR$, the Poincar\'e-Cartan form  of the Lagrangian $\lambda = h\rho$ takes the form
\beq\label{Cartan}
\theta_{\lambda} = \lambda - p_1 \cR(d\lambda) \,.
\eeq
In particular, $\theta_{h\rho}$ is a Lepage $n$-form; furthermore, by posing ${\pi^{r+1}_{r}}^* \rho = h\rho + \beta$, for a suitable order $s$, we have 
that $ 
{\pi_{r}}^* \rho - \theta_{h\rho} = \beta + p_1 \cR(dh\rho) \in \Theta^s_n$, with $\Theta^s_n\byd \ker p_n +d \ker p_{n-1}$.

\bRm \label{important remark}
If $\rho$  is a Lepage $n$-form then the dynamical form $p_1 d\rho = \cI(dh\rho)=\cI(d\rho)= E_{h\rho}$ is the {\em Euler--Lagrange  form} of the Lagrangian $h\rho$. 
\eRm

A  Lepage $n$-form can be seen as an {\em extension of a Lagrangian by a contact form};
thus, a Lepage equivalent of a given Lagrangian $\lambda$ is defined as a Lepage $n$-form $\rho$ such that $\lam=h\rho$ \cite{Kru73}.

It is well known that every Lagrangian admits a global Lepage equivalent. 
For $n=1$ and any $r$, or for any $n$ and $r \leq 2$ a global Lepage equivalent of $\lambda$ is the Poincar\'e-Cartan form.  
 For $n>1$, a Lepage equivalent of $\lambda$ is no longer unique (see in particular the discussion in \cite{Sau24}). 
 
\bRm
Apart from the (globalized) Poincar\'e-Cartan form, there are also other distinguished global Lepage equivalents of $\lambda$, among them the {\em Krupka--Betounes form} (for $r=1$) \cite{Betounes84,Kru77}, which has the property that $d\rho = 0$ if and only if $E_{h\rho} = 0$. 
\eRm

Given a Lagrangian $\lambda$ of order $r$, and a $\pi$-projectable vector field $\Xi$ on $Y$, it holds 
\beq \label{inf1vf}
L_{J^r\Xi}\lam \equiv L_{J^{2r}\Xi}h\rho= h(J^{2r-1}\Xi \rfloor
d\rho) + hd(J^{2r-1}\Xi \rfloor\rho) \equiv h L_{J^{2r-1}\Xi}\rho
\,,
\eeq
where $\rho$ is (any) Lepage equivalent of $\lambda$, and $L_{J^r\Xi}$ denotes the Lie derivative along the $r$-jet prolongation of $\Xi$.

\subsection{Recursive formulae and `full' Lepage equivalents} \label{Lep eq}

We saw that the Poincar\'e--Cartan equivalent  can be obtained {\em via} the residual operator of the differential of the Lagrangian \cite{PaRoWiMu16}.

Let us pose   
\beq
\rho_1=\lambda-p_1\mathcal{R}(d\lambda)=\theta_\lambda 
\eeq
with $\mathcal{R}$ the residual operator.
As we remarked we have the following remarkable property
\beq
p_1d \rho_1=p_1d \lambda- p_1d p_1\mathcal{R}(d\lambda)=p_1d\theta_\lambda = \cI(d \lambda )\,.
\eeq
In \cite{PaRoZa22} we obtained `full' Lepage equivalents generated by the Poincar\'e--Cartan form by the application of Rossi's recurrence formulae:
\bEq \label{recurrence formulae}
& & \rho_1=\lambda-p_1\mathcal{R}(d\lambda)   \nonumber \\
& &  \rho_2=\rho_1 - p_2\mathfrak{R}(d\rho_1)  \nonumber \\
& &  \rho_3=\rho_2 - p_3\mathfrak{R}(d\rho_2) \nonumber \\
& & \dots \nonumber \\
& &  \rho_n = \rho_{n-1}-p_n\mathfrak{R}(d\rho_{n-1}) \,,
\eEq
with $\mathfrak{R}$  the {\em lower} residual operator.

Note again that, as an extension of the above mentioned remarkable property, for $1\leq h \leq n$, the  Euler--Lagrange expressions of the $(h-1)$-contact equivalent of a Lagrangian are related with the $p_h$-contact component of the exterior differential  of the $(h-1)$-contact equivalent itself: 
\beq
p_h d \rho_h = p_h d \rho_{h-1}-p_h d p_h \mathfrak{R}(d\rho_{h-1}) = \mathfrak{I} (d\rho_{h-1}) \,.
\eeq

It is also noteworthy that such higher contact order Lepage equivalents are expressed by $\mathcal{R}$, and by the recursive use of the new operator $\mathfrak{R}$ defined in \cite{PaRoZa22}. The procedure above could be summarized introducing a sort of `long' residual operator by the formula $\rho_n = h\rho_{n} - \bar{R}(d\rho_{n})$ encompassing both of them.

In particular  we applied such recurrence formulae to obtain Lepage equivalents for first and second order Lagrangians, which are of particular interest in mathematical physics; the result can be summarized as follows \cite{PaRoZa22}:
\bPr\label{KBformula} 
 Let $\lambda 
 = \cL  ds$ be a Lagrangian on $J^r Y$.
\begin{itemize}

\item  the case $r=1$.

\beq \label{r=1}
\rho_n = \cL  ds+\sum_{q=1}^n\frac{1}{q!}\frac{\partial^q\cL }{\partial y^{\sigma_1}_{i_1}\dots\partial y^{\sigma_q}_{i_q}}\omega^{\sigma_1}\wedge\dots\wedge\omega^{\sigma_q}\wedge  \, ds_{i_1\dots i_q} \,,
\eeq
\ie the {\em Krupka--Betounes equivalent} of the Lagrangian $\lambda$;

\medskip

\item the case $r=2$.  Put   $f^{jm}_\sig:=p^{jm}_\sig  $, $f^j_\sig:=p^j_\sig - d_kp^{jk}_\sig$.

  \beq \label{2nd order extension}
 \rho_n =\cL  ds+\sum_{q=1}^n\frac{1}{q!}\frac{\der\cL }{\der y^{\sig_1}_{i_1}\dots\der y^{\sig_{q-1}}_{i_{q-1}}\der y^{\sig_q}_{i_qj}}\,\ome^{\sig_1}\wed\dots\wed\ome^{\sig_{q-1}}\wed\ome^{\sig_q}_j\wed ds_{i_1\dots i_q}+ \\
 + f^i_\sig\ome^\sig\wed ds_i +\sum_{q=1}^{n-1}\frac{1}{(q+1)!}\frac{\der f^{i_{q+1}}_{\sig_{q+1}}}{\der y^{\sig_1}_{i_1}\dots\der y^{\sig_q}_{i_q}}\,\ome^{\sig_1}\wed\dots\wed\ome^{\sig_q}\wed\ome^{\sig_{q+1}}\wed ds_{i_1\dots i_qi_{q+1}} \,;
\eeq
the latter reduces to the Krupka--Betounes equivalent when the Lagrangian is of order $r=1$. 

\end{itemize}
\ePr
 
\section{Applications: {\em two-fields} Lagrangians generating $2D$ modified and extended  Boussinesq equations} \label{3}

Let $n=3$, $m=2$, $r=2$, $Y=\mathbb{R}^2\times\mathbb{R}^3$, $\lam= \cL (x^i, y^\sig, y^\sig_{i_1}, y^\sig_{i_1 i_2}) ds$ be a Lagrangian on $J^2Y$.
For the sake of convenience, let us make the notational identifications
\beq
t  =x^1 , \quad
x =x^2, \quad
y  =x^3 , \quad
v  =y^1 , \quad
v_t =y^1_{1} , \quad
v_x =y^1_{2} , \quad
\eeq
\beq
w  =y^2 , \quad 
w_t =y^2_{1} , \quad 
w_x =y^2_{2} \,,
\eeq
and so on. We stress that here $v_t ,
v_x ,
w_t ,
w_x , \ldots $, {\em etc.} are local coordinates representing classes of sections of the fibered manifold which are equivalent up to second order partial derivatives at each fixed point and that they become  {\em partial derivatives only once they are pulled-back to the base manifold by holonomic sections} \cite{Sau89}.
With this notation, the formal  
derivatives are given by
\beq
& &  d_x =  \partial_x + v_x\partial_v + w_x\partial_w + v_{xt} \partial_{v_{t}} + v_{xy} \partial_{v_{y}} +  v_{xx} \partial_{v_{x}} + \\
& &  + w_{xt} \partial_{w_{t}} + w_{xy} \partial_{w_{y}} +  w_{xx} \partial_{w_{x}}+ \cdots \\
& &   d_y = \partial_y + v_y\partial_v + w_y\partial_w + v_{yt} \partial_{v_{t}} + v_{yy} \partial_{v_{y}} + v_{yx} \partial_{v_{x}} +\\
& &  + w_{yt} \partial_{w_{t}} + w_{yy} \partial_{w_{y}} +  w_{yx} \partial_{w_{x}}+ \cdots \\
& &   d_t =  \partial_t + v_t\partial_v + w_t\partial_w + v_{tt} \partial_{v_{t}} + v_{ty} \partial_{v_{y}} + v_{tx} \partial_{v_{x}} + \\
& &   + w_{tt} \partial_{w_{t}} + w_{ty} \partial_{w_{y}} +  w_{tx} \partial_{w_{x}}+ \cdots 
\eeq
Note that, according to our general definitions of the notation, for convenience, sometimes we shall denote  by $ds_{tx}$ the differential $dy$, or by $ds_t$ the exterior product $dx\wedge dy$, and so on (taking into account that $t, x, y$ are in fact ordered, since they correspond to $x^1, x^2, x^3$). Furthermore, in the following by a slight abuse of notation, the index $i$ will run over $t,x,y$ and the index $\sigma$ will run over $v, w$. Accordingly by the compact notation $\partial^i_v$ we shall denote $\frac{\partial}{\partial v_i}$, \eg $\partial^2_v =\frac{\partial}{\partial v_2}=\frac{\partial}{\partial v_x}$ and a coherent notation holds true for the local expression of contact $1$-forms.

In what follows, we consider some two-field Lagrangians, construct their Lepage equivalents of Krupka-Betounes type and obtain various Boussinesq models by taking the first degree contact component of the exterior differential of such equivalents, say $p_{1}d\rho$. The results automatically characterize the Boussinesq models generated by this procedure as variational (\ie as E-L equations). We should notice that the Krupka-Betounes equivalents of null Lagrangians (\ie divergences) are closed under the exterior differential \cite{Betounes84, Betounes98,Kru73,Kru93}. 

Note that, although the E-L equations remain unchanged by adding to the Lagrangian a divergence, Krupka-Betounes equivalents in general define different conserved currents, depending on which divergence is added to the Lagrangian.

\subsection{Fourth-order $(2+1)$ extensions with mixed spatial derivatives} \label{general case}
 
Let $\lambda=\cL ds $ be a Lagrangian form on $J^2 Y$, where  
 the Lagrangian density $\cL $ (by a slight abuse also called in the following `Lagrangian' ) is given by
\bEq  \label{Lagrangian 1 constrained}
\cL = 
w^2 +\frac{1}{2} v_x^2
 + a w_t v_{xx} 
 - \frac{1}{2}a v_{tt} v_{xx}
 -\frac{1}{2}b v_{xx}^2+\frac{1}{3} v_x^3 + \frac{1}{2}\beta  v_{y}^2 + \gamma (w - v_{t}) 
\eEq
where $\gamma (t,x,y) $ plays the role of a Lagrange multiplier. 

By applying Proposition \ref{KBformula},  in the case $r=2$ we obtain the following local description of the `full' equivalent associated with the Lagrangian above.

\bPr
Let $\theta_\lambda$ be the Poincaré-Cartan form associated with the Lagrangian form $\lambda=\cL ds$.
The `full'  extended Lepage equivalent for Lagrangian  \eqref{Lagrangian 1 constrained} is given by
\bEq \label{main Lepage}   
   \rho_3
= \theta_\lambda+\frac{1}{2}a\omega^w\wedge\omega^v_x\wedge ds_{tx} =  \theta_\lambda+\frac{1}{2}a \omega^w \wedge \omega^v_x \wedge dy \,.
\eEq
\ePr

\bPf
In the case $r=2$ and for $n=3$  Proposition \ref{KBformula} reads
\bEq \label{main Lepage2}
        \rho_3 & = \cL ds+\sum_{q=1}^{3}\frac{1}{q!}\frac{\partial\cL}{\partial y^{\sigma_1}_{i_1}\cdots\partial y^{\sigma_{q-1}}_{i_{q-1}}\partial y^{\sigma_q}_{i_qj}}\omega^{\sigma_1}\wedge\cdots\wedge\omega^{\sigma_{q-1}}\wedge\omega^{\sigma_q}_j\wedge ds_{i_1\cdots i_q} \\
        &\quad+f^i_\sigma\omega^\sigma\wedge ds_i+\sum_{q=1}^{2}\frac{1}{(q+1)!}\frac{\partial f^{i_{q+1}}_{\sigma_{q+1}}}{\partial y^{\sigma_1}_{i_1}\cdots\partial y^{\sigma_q}_{i_q}}\omega^{\sigma_1}\wedge\cdots\wedge\omega^{\sigma_{q+1}}\wedge ds_{i_1\cdots i_{q+1}} \,, \nonumber
\eEq
where $f^i_\sigma=p^i_\sigma-d_kp^{ik}_\sigma$. 

In order to find the explicit expression of $\rho_3$, we first need to compute the quantities:
\beq
& & f^t_v = p^t_v-d_kp^{tk}_v = -  \gamma + \frac{1}{2} av_{txx}, \quad f^t_w = av_{xx},  \quad \\
& & f^x_v = v_x+v_x^2 - aw_{tx} + \frac{1}{2}av_{ttx}+bv_{xxx},    \quad  f^y_v = \beta v_y, \quad \\
& & f^x_w = f^y_w = 0, \quad  p^{tt}_v = - \frac{1}{2}av_{xx}, \quad  p^{xx}_v = aw_t-\frac{1}{2}av_{tt}-bv_{xx},\\
& & p^{tx}_v = p^{ty}_v = p^{xy}_v = p^{yy}_v = p^{\cdot\cdot}_w=0 \,,
\eeq
with obvious meaning of dotted notation.

The first summation in \eqref{main Lepage2} splits as follows.
To the value $q=1$ corresponds the usual component defining the Poincar\'e-Cartan extension, so that as a byproduct, we have associated with the $2D$ Boussinesq model above the explicit expression for the Poincar\'e-Cartan equivalent, well known to define momentum map and Hamiltonian content of the model itself,
\beq
-\frac{1}{2}av_{xx}\omega^v_t\wedge ds_t + (aw_t-\frac{1}{2}av_{tt}-bv_{xx})\omega^v_x\wedge ds_x \, ;
\eeq
while the value $q=2$ provides an additional term to the Poincar\'e-Cartan equivalent:
\beq
\frac{1}{2}[\partial( aw_t-\frac{1}{2}av_{tt}-bv_{xx} )/\partial w_t ] \omega^w\wedge\omega^v_x\wedge ds_{tx} =\frac{1}{2}a \omega^w\wedge\omega^v_x\wedge ds_{tx} \,.
\eeq
We also have, with an obvious meaning of the notation,
$\partial ( av_{xx})/ \partial y^{\sigma_1}_{i_1}\cdots\partial y^{\sigma_{q-1}}_{i_{q-1}} = 0$.
Concerning the second summation in \eqref{main Lepage2}, the only terms that could in principle contribute are
$\frac{1}{2}\frac{\partial f^x_v}{\partial v_x}$ and $\frac{1}{2}\frac{\partial f^y_v}{\partial v_y}$,
but both give rise to vanishing terms because they multiply, respectively, the forms
$\omega^v\wedge\omega^v\wedge ds_{xx}=0$ and $\omega^v\wedge\omega^v\wedge ds_{yy} = 0$ (this is a consequence of the particular expression of the Lagrangian).

Collecting all such considerations,  we finally get the following expression for the `full' equivalent of  $\lambda$
\bEq   \label{full equivalent}
  & &  \rho_3= \cL ds+ ( \frac{1}{2}av_{txx}- \gamma ) \omega^v\wedge ds_t+  \\ 
    & &  (av_{xx})\omega^w\wedge ds_t  
  + ( v_x+v_x^2-aw_{tx}+\frac{1}{2}av_{ttx}+bv_{xxx} ) \omega^v\wedge ds_x+  \nonumber \\ 
   & &  (\beta v_y)\omega^v\wedge ds_y  + ( -\frac{1}{2}av_{xx} ) \omega^v_t\wedge ds_t+ ( aw_t-\frac{1}{2}av_{tt}-bv_{xx} ) \omega^v_x\wedge ds_x +   \nonumber \\ 
   & &  \frac{1}{2}a\omega^w\wedge\omega^v_x\wedge ds_{tx} 
= \theta_\lambda+\frac{1}{2}a\omega^w\wedge\omega^v_x\wedge ds_{tx} \equiv \theta_\lambda+\frac{1}{2}a \omega^w \wedge \omega^v_x \wedge dy \,, \nonumber
\eEq
with $\theta_\lambda$ the Poincar\'e-Cartan equivalent of $\lambda$.
\ePf\begin{flushright}
$\Box$
\end{flushright}
According to Remark \ref{important remark}, in a local chart on $J^4Y$, the corresponding E-L form is given by $p_1 d \rho_3  = \cI (d\lambda)$:
\beq
& &    p_1d\rho_3 = \\
& &  =  ( 2w\omega^w+v_x\omega^v_x+av_{xx}\omega^w_t+aw_t\omega^v_{xx}-\frac{1}{2}av_{xx}\omega^v_{tt}-\frac{1}{2}av_{tt}\omega^v_{xx}-bv_{xx}\omega^v_{xx}    +v^2_x\omega^v_x  \\
& & +\beta v_y\omega^v_y + {\gamma\omega^w}-\gamma\omega^v_t )\wedge ds+ \frac{1}{2}av_{xx} \omega^v_{tt}\wedge ds   - ( aw_t-\frac{1}{2}av_{tt}-bv_{xx}  ) \omega^v_{xx}\wedge ds \\
& & - (\frac{1}{2}av_{txx}-\gamma) \omega^v_{t}\wedge ds   - ( v_x+v_x^2-aw_{tx}+\frac{1}{2}av_{ttx}+bv_{xxx}  ) \omega^v_{x}\wedge ds - \beta v_y \omega^v_y\wedge ds  \\
& & +av_{xx} ( -\omega^w_t\wedge ds  )+ ( \frac{1}{2}av_{txx}  )\omega^v_t\wedge ds+ ( -aw_{tx}+\frac{1}{2}av_{ttx}+bv_{xxx}  )\omega^v_x\wedge ds \\
& &  -\frac{1}{2}av_{ttxx} \omega^v\wedge ds + ( {aw_{txx}-v_{xx}-2v_xv_{xx}-\frac{1}{2}av_{ttxx}-bv_{xxxx}} +\gamma_t )\omega^v\wedge ds  \\
& & -\beta v_{yy}\omega^v\wedge ds - av_{txx}\omega^w\wedge ds  = (2w+\gamma-av_{txx})\omega^w\wedge ds \\
& &  +(aw_{txx}-v_{xx}-2v_xv_{xx}-av_{ttxx}-bv_{xxxx}  -\beta v_{yy} + \gamma_t)\omega^v\wedge ds \,.
\eeq
Now, the E-L equations are given by $p_1d\rho_3\circ j^4  \sigma=0$  and they split as  
\bEq 
& & v_{xx}+2v_xv_{xx} - a w_{txx}+bv_{xxxx}+\beta v_{yy}+av_{ttxx}   - \gamma_t   = 0 \,, \\
& & 2w -  a v_{txx}  +\gamma  = 0 \,,
\eEq
with the constraint 
\bEq \label{constraint1}
w=v_t \,.
\eEq
The first two provide us the equation
\bEq 
v_{xx}+2v_xv_{xx} - a w_{txx}+bv_{xxxx}+\beta v_{yy}+av_{ttxx}  - 2w_t  +a v_{ttxx} = 0 \,,
\eEq
which with the constraint \eqref{constraint1} becomes the following (2+1)-dimensional  Boussinesq equation
\bEq 
v_{xx}+2v_xv_{xx} + a v_{ttxx}+bv_{xxxx}+\beta v_{yy} + 2v_{tt}  =  0 \,.
\eEq

\bRm
For $\beta=-1$ and letting $y$ play the role of time and $t$ the role of a spatial variable, this equation appears as a modification and an extension in $(2+1)$ dimensions of the modified Boussinesq equation in potential form. Apparently it is resembling a particular case of \cite{RGRB19}, however, note that here the mixed term is involving derivatives of the two space variables.
In fact, it could be rather seen  as an extension of  certain models describing the water problem in $(1+1)$ dimensions with surface tension \cite{ScWa01},
or mechanical waves in myelinated axons \cite{TPE22}.
\eRm

\subsection{Two-field derivation of fourth-order $(2+1)$ extensions}

Let now consider the following Lagrangian
\bEq  \label{Lagrangian 3 constrained}
\cL = w^2 +\frac{1}{2} v_x^2 + a w_t v_{xx}  - \frac{1}{2}a v_{tt} v_{xx} - \frac{1}{2}b v_{xx}^2+\frac{1}{3} v_x^3 + \frac{1}{2}\beta  v_{y}^2 + \gamma (w - \frac{1}{2}v_{t})
\eEq
describing a $2$-field variational Boussinesq type model along a different 
constraint.

We can proceed as above in order to get the E-L equations
\bEq 
v_{xx}+2v_xv_{xx} - a w_{txx}+bv_{xxxx}+\beta v_{yy}+av_{ttxx}  -\frac{1}{2}\gamma_t     = 0 \,;
\eEq
\bEq
2 w -   a v_{txx}  +  \gamma  = 0 \,,
\eEq
with  the constraint equation
\bEq \label{constraint}
w = \frac{1}{2} v_t \,.
\eEq
Now we get $ \gamma_t = a v_{ttxx} -2w_t$ and substituting in the first E-L equation we have

\bEq 
v_{xx}+2v_x v_{xx} - a w_{txx}+bv_{xxxx}+\beta v_{yy}   +    a v_{ttxx}   - \frac{1}{2} a v_{ttxx}  + w_t   = 0 \,;
\eEq
which with the constraint equation gives us 
\bEq 
v_{xx}+2v_x v_{xx} +bv_{xxxx}+\beta v_{yy}   + v_{tt}   = 0 \,.
\eEq

\bRm
Again putting $\beta=-1$ and interpreting the variable $t$ as a spatial variable and the variable $y$ as the time, this equation resembles (up to coefficients) a particular case of 
the $(2+1)$ variational Boussinesq equation derived as a single E-L equation from a $1$-field Lagrangian in \cite{AGR18}.
\eRm

We can consider several different constrained Lagrangian systems and obtain many different kinds of Boussinesq equations. The one which follows is particularly interesting.
Let the Lagrangian be given as
\beq
 \lambda&=\left(\frac{1}{2}w^2+\frac{1}{2}v_x^2+aw_tv_{xx}-\frac{1}{2}av_{tt}v_{xx}-\frac{1}{2}bv_{xx}^2+\frac{1}{3}v_{x}^3+\frac{1}{2}\beta v_{y}^2+\gamma(w+v_t)\right)ds,
\eeq
and again by a similar procedure as above, we get
\beq
& & v_{xx}-aw_{txx}+av_{ttxx}+bv_{xxxx}+2v_xv_{xx}+\beta v_{yy}+\gamma_t=0 \\
& & w-av_{txx}+\gamma=0,
\eeq
with constraint equation $w+v_t=0$
and, after some manipulations, the equation
\beq
v_{xx}-aw_{txx}+av_{ttxx}+bv_{xxxx}+2v_xv_{xx}+\beta v_{yy}- av_{ttxx}+w_t = 0 \,,
\eeq
by means of the constraint, provide a {\em fourth-order mixed term} $(2+1)$ Boussinesq type equation
\beq
v_{xx}+av_{ttxx}+bv_{xxxx}+2v_xv_{xx}+\beta v_{yy}-v_{tt} = 0 \,.
\eeq

\bRm
This equation appears as a modification and an extension in $(2+1)$ dimensions of the Boussinesq equation in potential form. Indeed, it can be recognized as a particular case   (for $f(v_x )= v_x^2$) of \cite{RGRB19}.
\eRm

\subsection{Sixth-order $(2+1)$ extensions with mixed spatial derivatives} \label{boh?}

Let now $\gamma = 0$ (non-constrained case)
and let us then determine the corresponding  Lepage equivalent and equations. As for the Lagrangian we have 
\bEq \label{Lagrangian 2}
\mathcal{L}=w^2  + \frac{1}{2}v_x^2+ aw_tv_{xx}-\frac{1}{2}av_{tt}v_{xx}-\frac{1}{2}bv_{xx}^2+\frac{1}{3}v_x^3+\frac{1}{2}\beta v_y^2  \,,
\eEq
and the `full' extended Lepage equivalent for Lagrangian \eqref{Lagrangian 2} is given by
\beq
\rho_3 = \theta_\lambda+\frac{1}{2}a\omega^w\wedge\omega^v_x\wedge ds_{tx} =  \theta_\lambda+\frac{1}{2}a \omega^w \wedge \omega^v_x \wedge dy \,.
\eeq

Again according to Remark \ref{important remark}, the Euler-Lagrange equations are given by a relation of the kind $p_{1}d \rho_3 = \cI (d\lambda)$ adapted to this case. As in the previous example, they split as  
\bEq\label{equazione da studiare}
v_{xx} + 2 v_{x}v_{xx} + a v_{ttxx} - a w_{txx} + b v_{xxxx} + \beta v_{yy} = 0 \,,
\eEq
\bEq
w - \frac{1}{2} a v_{xxt}=0 \,,
\eEq
from which we get 
\bEq \label{most general Boussinesq}
v_{xx}+2v_xv_{xx}-\frac{1}{2}a^2v_{ttxxxx}+bv_{xxxx}+\beta v_{yy}+av_{ttxx}=0 \,;
\eEq
as above,  by setting $\beta=-1$, and letting $y$ play the role of a time and $t$ of a spatial coordinate, we obtain the following {\em sixth-order} $(2+1)$ Boussinesq-type equation
\bEq
v_{yy}=v_{xx}+2v_xv_{xx}-\frac{1}{2}a^2v_{ttxxxx}+bv_{xxxx}+av_{ttxx} \,.
\eEq

\bRm
The equation above is of Boussinesq type again extending the ones in  
\cite{ScWa01,TPE22}.

It also {\em resembles} sixth-order mixed derivative which are extensions of fourth order mixed terms dispersion $(1+1)$-dimensional Boussinesq equations,
see \eg \cite{RGRB19} and references therein. 
Note however that here again {\em the mixed term actually involves two spatial coordinates} 
and not one time and one spatial coordinate, 
and it appears within a $(2+1)$ Boussinesq equation, extending to higher orders the mixed  terms added in \cite{WaWaMa18,WaKa19,ZhMaMc21}.
The $(2+1)$ extension of the $(1+1)$ model keeps account of the sixth-order {\em spatial} mixed derivatives contribution and it is not 
just the addition of a second order derivative in the other spatial coordinate.
\eRm 

\subsubsection{The case of the reduced KP equation in the moving-frame}
For $a=0$ the latter equation is a $(1+1)$-dimensional Boussinesq equation given in potential form (here $y$ playing the role of time). In fact, it corresponds to the `potential form' of the original Boussinesq equation seen as a reduction (searching for stationary solutions) of the so-called $(2+1)$-dimensional KP equation taken in the moving frame (according to \cite{BoZa02})\footnote{Note that in \cite{KaRo23} a KP equation in the moving frame is defined as the original KP equation: of course, the concepts of fixed and moving frame are of relative nature.}. 
Thus we have obtained a {\em $2$-field variational characterization} of the reduced moving-frame KP equation given in potential form. 

\section{Symmetries and `improved' conserved currents}

The question arises now whether there exist symmetries (among the symmetry transformations of extremals) which are symmetries of those Lepage equivalents, but neither symmetries of the Lagrangians nor of the Poincaré-Cartan equivalent.
In the following, we show that if such symmetries exist, then it is possible to prove the existence (and the expression) of `improved' Noether currents associated with them.
In what follows, a subscript to $\rho$ indicates the Lagrangian of which we are considering the Lepage equivalent. 

First of all, we recall that for first order field theories the Krupka-Betounes equivalent satisfies the so-called mapping property, \ie the property $L_{J^{1} \Xi} \rho_\lambda = \rho_{ L_{J^{1} \Xi} \lambda}$ holds true. 
In the case of first order field theories, relaying on such a property, Betounes proved the existence of an `improved' conservation law associated with invariance symmetries of the `full' Lepage equivalent \cite{Betounes98}.
As well known, a way to find out such symmetries is to look for projectable vector fields such that $\cE_n ( L_{J^r \Xi}\lambda) =0$ {\em along extremals} (so-called point symmetries). Notice that these are in fact Jacobi fields along extremals, see \cite{AcPa21} Th. $IV.3$ and Th. $IV.5.$

In the following, for any fixed order $r$, we investigate the existence of `improved' conservation laws associated with invariance with respect to symmetry transformations of the set of extremals.
\bTh \label{THEOREM}
For any projectable vector field $\Xi$ such that $\cE_n ( L_{J^r \Xi}\lambda) = 0$ along any critical section, we have
 \bEq 
 d_H(\Xi\rfloor \rho_\lambda - \psi) = 0\,,
 \eEq
where $\rho_\lambda$ is the `full' Lepage equivalent of $\lambda$ given by the recursive formulae \eqref{recurrence formulae} and $\psi$ is defined up to a horizontal differential by the relation $d_H  \psi = h \rho_{ L_{J^{2r+1} \Xi}\lambda}$ \,.
\eTh

\bPf
Let us explicate the condition $\cE_n ( L_{J^r \Xi}\lambda) =0$ in terms of the Lepage equivalent $\rho_\lambda$.

Because of the naturality of the variational Lie derivative we have 
\beq
 \cE_n ( L_{J^r \Xi}\lambda)  =  L_{J^{2r+1} \Xi} \cE_n (\lambda) \,;
\eeq
on the other hand, recall that for any global Lepage equivalent 
$\rho_\lambda$ of $\lambda$, we have
\beq
  L_{J^{2r+1} \Xi} \cE_n (\lambda) =  L_{J^{2r+1} \Xi} p_1d \rho_\lambda \,.
 \eeq

Note that, since the Lie derivative with respect to projectable vector fields preserves the degree of contactness \cite{Kru73}, we have
\beq
 L_{J^{2r+1} \Xi} p_1d \rho_\lambda  = p_1   L_{J^{2r+1} \Xi} d \rho_\lambda \,,
\eeq
and again by naturality
\beq
p_1   L_{J^{2r+1} \Xi} d \rho_\lambda = p_1 d   L_{J^{2r+1} \Xi} \rho_\lambda  \,.
\eeq
Now, in the following we prove that $ L_{J^{2r+1} \Xi} \rho_\lambda$ is again a Lepage equivalent of  a Lagrangian.

Let us then consider the identity
\bEq
 L_{J^{2r+2} \Xi} \lambda = L_{J^{2r+2} \Xi} \lambda 
\eEq
and, by the very definition of a Lepage equivalent and since  $L_{J^{2r+1} \Xi}$ preserves the contact splitting, so that
$h   L_{J^{2r+1} \Xi} \rho_\lambda = L_{J^{2r+2} \Xi} h  \rho_\lambda =  L_{J^{2r+2} \Xi} \lambda $,
 write it as 
\bEq
h \rho_{ L_{J^{2r+1} \Xi} \lambda} =  h  L_{J^{2r+1} \Xi}\rho_{\lambda} \,,
\eEq
which implies
\bEq
\cE_n (h \rho_{ L_{J^{2r+1} \Xi} \lambda}) = \cE_n ( h  L_{J^{2r+1} \Xi}\rho_{\lambda}) \,.
\eEq
Now, on the one hand we have
 \beq
 \cE_n (  L_{J^{2r+1} \Xi} \lambda) =\cE_n (h \rho_{ L_{J^{2r+1} \Xi} \lambda})= p_1d \rho_{ L_{J^{2r+1} \Xi} \lambda} \,,
 \eeq
while on the other hand
\beq
\cE_n (  L_{J^{2r+1} \Xi} \lambda) =\cE_n (  L_{J^{2r+1} \Xi}h \rho_\lambda)=   L_{J^{2r+1} \Xi} \cE_n (h  \rho_\lambda) = L_{J^{2r+1} \Xi}  p_1 d   \rho_\lambda =p_1d  L_{J^{2r+1} \Xi} \rho_\lambda\,,
\eeq
so that 
\bEq \label{weak mapping}
p_1d \rho_{ L_{J^{2r+1} \Xi} \lambda}=  p_1 d   L_{J^{2r+1} \Xi} \rho_\lambda \,.
\eEq

An interesting question is whether such a Lepage equivalent is exactly the one associated with the Lie derivative of the original Lagrangian $\lambda$ (the so-called mapping property). 
The condition above is satisfied in particular if
\bEq
\rho_{ L_{J^{2r+1} \Xi} \lambda} =  L_{J^{2r+1} \Xi} \rho_\lambda + p_1d \eta \,,
\eEq
\ie the mapping property 
surely holds true up to the $1$-contact part of a locally exact $n$-form.

Also the question arises whether such a Lepage equivalent is a Krupka-Betounes equivalent, \ie, if $d \rho_{ L_{J^{2r+1} \Xi} \lambda}=0$ implies $\cE_n(L_{J^{2r+1} \Xi} \lambda)=0$ , but  this is the case if we assume that $\Xi$ is a symmetry of the Euler--Lagrange equations, and a similar property holds true along extremals if $\Xi$ is a symmetry transformation of extremals.

In any case, we note that, for our purposes, only a kind of mapping property is required to hold true for the $0$-contact (horizontal) components.
Indeed, by the definition of a Lepage equivalent, our result guarantees that $ L_{J^{2r+1} \Xi} \rho_\lambda$ as a Lepage equivalent can be identified  with $\rho_{ L_{J^{2r+1} \Xi} \lambda}$, \ie it is a Lepage equivalent of  the Lagrangian $h L_{J^{2r+1} \Xi} \rho_\lambda = L_{J^{2r+1} \Xi} \lambda$.

From \eqref{weak mapping} we also get
$\cE_n (h   L_{J^{2r+1} \Xi} \rho_\lambda) = \cE_n (h  \rho_{ L_{J^{2r+1} \Xi} \lambda}) $, so that apparently, $h  L_{J^{2r+1} \Xi} \rho_\lambda =  h  \rho_{ L_{J^{2r+1} \Xi} \lambda} +d_H\nu$, but it is easy to see that $d_H\nu=0$, as we already noted that,
for any projectable vector field being a symmetry of the equations, the identity $h L_{J^{2r+1} \Xi} \rho_\lambda =  h  \rho_{ L_{J^{2r+1} \Xi} \lambda}$ holds true.

Now, on the one hand the condition $\cE_n (h    L_{J^{2r+1} \Xi} \rho_\lambda) =0$ means
that locally 
\beq
h    L_{J^{2r+1} \Xi} \rho_\lambda = d_H \gamma \,,
\eeq
on the other hand, from
 $\cE_n (h   \rho_{ L_{J^{2r+1} \Xi} \lambda} )=0$, we have locally
\bEq \label{Betounes}
h  \rho_{ L_{J^{2r+1} \Xi} \lambda} = d_H \psi \,.
\eEq 
We then get the following extension -for higher order Lagrangians- of the analogous Betounes' result for first order field theories
\bEq  \label{IMPORTANTE}
d_H \gamma = h  L_{J^{2r+1} \Xi} \rho_\lambda =  h  (\Xi \rfloor d\rho_\lambda +d(  \Xi \rfloor \rho_\lambda) )  =  h   \rho_{ L_{J^{2r+1} \Xi} \lambda } = d_H  \psi  \,.
 \eEq
In particular,
\bEq\label{fundamental}
\Xi_V \rfloor p_1d \rho_\lambda +d_H(\Xi\rfloor \rho_\lambda)=  d_H \psi\,,
\eEq 
where $\psi=\gamma+d_H\mu$,
and, along any critical section such that $p_1d \rho_\lambda = \cE_n(\lambda) =0$, (pull-back by prolongantion of critical sections omitted to simplify notation) we have the {\em `weak' conservation law}
 \bEq 
 d_H(\Xi\rfloor \rho_\lambda - \psi) = 0\,,
 \eEq
which is the desired higher order extension of a result obtained by Betounes in first order theories \cite{Betounes98}.
\ePf
\begin{flushright}
$\Box$
\end{flushright}

\bRm
 It is noteworthy that \eqref{fundamental} generalizes the Noether Theorem II \cite{Noe18}. Indeed, from equation \eqref{fundamental} by posing $\epsilon =\Xi\rfloor \rho_\lambda -\psi$, we get the following reformulation of the Noether Theorem II, holding along {\em any }generic holonomic section
\bEq \label{fundamental2}
d_H\epsilon = - \Xi_V \rfloor p_1d \rho_\lambda \,,
\eEq
which also  extends the analogous Betounes' formula for first order theories \cite{Betounes98}. 

Of course, given $\psi$, the above can be interpreted as an equation for the symmetry $\Xi$ and generalizes the classical Noether--Bessel-Hagen equation.
\eRm

\bRm
It is well known that for gauge-natural theories, it is easy to manipulate the term $\Xi_V \rfloor p_1d \rho_\lambda$ in order to split it as the sum of Noether identities plus a divergence (a horizontal differential in our framework). In particular, one uses linearity properties of the vertical part of the gauge-natural lift of principal infinitesimal automorphisms,
 and its relationship with the Lie derivative of sections of the gauge-natural bundle, and 
performs an integration by parts, see \eg \cite{FFP01,FFPW11} for more detail. Although we do not have such a rich structure on the bundle of configurations, in the hypotheses of Theorem \ref{THEOREM}, formula \eqref{fundamental2} gives us a suitable generalization of the Noether Theorem II. Indeed, $\Xi_V$ is constrained by the request to be generators of symmetry transformations of extremals, \ie solutions of the Jacobi equations
 \cite{AcPa21}, providing a link between $\Xi^\alpha$ and $\xi^i$ and their derivatives. By integration by parts on the derivatives of $\xi^i$,
an explicit expression of $\Xi_V \rfloor p_1d \rho_\lambda$ as a full horizontal differential could be obtained. 
\eRm

\bEx \label{esempio corrente conservata}
We can calculate explicitly $d_H \psi$  alternatively from \eqref{Betounes} or using its characterization  \eqref{fundamental}.
As for the first possibility we can proceed as follows.

Let $\Xi$ be a {\em projectable vector field} on the fibration $\mathbb{R}^5 \to \mathbb{R}^3\colon(t,x,y,v,w)\mapsto(t,x,y)$ and let us denote by $J^1\Xi$ its first order prolongation (for the prolongation of projectable vector fields, see \eg \cite{Sau89}) given in local fibered coordinates by 
\bEq \label{jet prol}
& &  J^1\Xi = \Xi^v_{t}\frac{\partial}{\partial v_{t}}+\Xi^v_{x}\frac{\partial}{\partial v_{x}}+\Xi^v_{y}\frac{\partial}{\partial v_{y}}+\Xi^w_{t}\frac{\partial}{\partial w_{t}}+\Xi^w_{x}\frac{\partial}{\partial w_{x}}+\Xi^w_{y}\frac{\partial}{\partial w_{y}}  \\ \nonumber
& &  +\Xi^v\frac{\partial}{\partial v}+\Xi^w\frac{\partial}{\partial w}+ \xi^t\frac{\partial}{\partial t}+\xi^x\frac{\partial}{\partial x}+\xi^y\frac{\partial}{\partial y}  \,.
\eEq
Here the subscripts in $\Xi$ merely denote labels and {\em are not} total derivatives in general (unless the vector field is vertical); their explicit expression can be found \eg in \cite{Kru73,Sau89}.

Let $\lambda = \mathcal{L}ds$ be  the Lagrangian \eqref{Lagrangian 2} given in Subsection \ref{boh?} and let us denote
\beq
L_{J^2\Xi}\lambda = \Xi_V \rfloor \cI d (\lambda) + d_H (J^{r-1}\Xi_V \rfloor p_{d_{V }\lambda} + \Xi_ H \rfloor \lambda) = \mathcal{L}'ds \,,
\eeq
with $p_{d_{V }\lambda}=  - p_1\mathcal{R}(d \lambda) \equiv \theta_\lambda - \lambda$,
where $\mathcal{L}' $ is the  new Lagrangian density given by
\beq
& & \mathcal{L}'  =  \partial_i\xi^i\mathcal{L}+2w\Xi^w+(v_x+v_x^2)\Xi^v_x+\beta v_y\Xi^v_y+av_{xx}\Xi^w_t 
+  \\
& & + (aw_t-\frac{1}{2}av_{tt}-bv_{xx})\Xi^v_{xx}-\frac{1}{2}av_{xx}\Xi^v_{tt}   \,. 
\eeq
The `full' equivalent $\rho_{L_{J^2\Xi}\lambda} $ appearing  in \eqref{Betounes}, 
is given by Proposition \ref{KBformula}, for $r=2$, \ie 
\beq 
& & \rho_{L_{J^2\Xi}\lambda} = 
\mathcal{L}'ds+\frac{\partial\mathcal{L}'}{\partial y^{\sigma_1}_{i_1j}}\omega^{\sigma_1}_j\wedge ds_{i_1}+\frac{1}{2}\frac{\partial\mathcal{L}'}{\partial y^{\sigma_1}_{i_1}\partial y^{\sigma_2}_{i_2j}}\omega^{\sigma_1}\wedge\omega^{\sigma_2}_j\wedge ds_{i_1i_2} + \\ \nonumber 
& & \frac{1}{6}\frac{\partial\mathcal{L}'}{\partial y^{\sigma_1}_{i_1}\partial y^{\sigma_2}_{i_2}\partial y^{\sigma_3}_{i_3j}}\omega^{\sigma_1}\wedge\omega^{\sigma_2}\wedge\omega^{\sigma_3}_j\wedge ds_{i_1i_2i_3}  +
\\ \nonumber 
& & 
f^i_\sigma\omega^\sigma\wedge ds_i +\frac{1}{2}\frac{\partial f^{i_2}_{\sigma_2}}{\partial y^{\sigma_1}_{i_1}}\omega^{\sigma_1}\wedge\omega^{\sigma_2}\wedge ds_{i_1i_2} \,.
\eeq
We note, however, that its contact components do not play a role in the computation of $h \rho_{ L_{J^{2r+1} \Xi} \lambda}$ being  $h \rho_{ L_{J^{2r+1} \Xi} \lambda}= \cL ' ds$; nonetheless, their explicit expression can be of interest for a direct comparison with the approach by Betounes, see \eg \cite{PaZa23}.

The vertical part of the projectable vector field $\Xi=\xi^i\partial_i+\Xi^\alpha\partial_\alpha$ on $Y$ is a generalized vertical vector field $\Xi_V=\Xi_V^\alpha\partial_\alpha$ on $Y$ whose components are (locally) given by $\Xi_V^\alpha=\Xi^\alpha - y^\alpha_j \xi^j $, and for its first prolongation ${\Xi_V}^\alpha_i=\Xi^\alpha_i - y^\alpha_{j i }\xi^j$.

We therefore get 
\bEq\label{first expression}
& & d_H  \psi= \cL ' ds    = \Xi_V \rfloor \cE_3 (\lambda) + d_H (J^{1}\Xi_V \rfloor p_{d_{V }\lambda} + \Xi_ H \rfloor \lambda)  =
\\ \nonumber
& & 
 \Xi_V \rfloor \cE_3 (\lambda)  + d_H (  J^{1}\Xi_V \rfloor (  \theta_\lambda -\lambda)    + \Xi_ H \rfloor \lambda  ) =
 \\ \nonumber
& &
 \Xi_V \rfloor \cE_3 (\lambda) +  d_H [(\Xi^\alpha - y^\alpha_j \xi^j ) \partial_\alpha + (\Xi^\alpha_i - y^\alpha_{j i}\xi^j ) \partial^i_\alpha \rfloor  (  \frac{1}{2}av_{txx}
\, \,  \omega^v\wedge ds_t +  \\  \nonumber
    & &  av_{xx} \, \,  \omega^w\wedge ds_t  
  + ( v_x+v_x^2-aw_{tx}+\frac{1}{2}av_{ttx}+bv_{xxx} ) \omega^v\wedge ds_x+ \\  \nonumber 
   & &  \beta v_y \, \,  \omega^v\wedge ds_y   -\frac{1}{2}av_{xx} \, \,  \omega^v_t\wedge ds_t+ ( aw_t-\frac{1}{2}av_{tt}-bv_{xx} ) \omega^v_x\wedge ds_x) +
   ( \xi^id_i )\, \, \rfloor
    \\ \nonumber
& &
 (w^2 +\frac{1}{2} v_x^2   + a w_t v_{xx} 
 - \frac{1}{2}a v_{tt} v_{xx}
 -\frac{1}{2}b v_{xx}^2+\frac{1}{3} v_x^3 + \frac{1}{2}\beta  v_{y}^2  )ds]\,.
\eEq

We can compare this result with the one coming from the alternative path.
In fact, we saw that an implicit local expression of the current $\psi$ is also given by  
\bEq \label{boh}
 d_H  \psi  = \Xi_V \rfloor p_1d \rho_3 +  d_H ( J^{1}\Xi\rfloor \rho_3)  = \Xi_V \rfloor \cE_3 (\lambda) + d_H ( J^{1}\Xi\rfloor \rho_3)  \,,
\eEq
where $\rho_3\equiv ({\rho_\lambda})_3$ is given by equation \eqref{full equivalent}, for $\gamma=0$. For the case of study it gives the local coordinate expressions
\bEq  \label{straboh}
& & d_H \psi = \Xi_V \rfloor \cE_3 (\lambda) + d_H [(\xi^i(\partial_i+v_i\partial_v+ w_i \partial_w  + v_{ij}\partial_v^j + w_{ij } \partial_w^j )+\\ \nonumber
& & ( (\Xi^v-v_j \xi^j) \partial_v +  (\Xi^w-w_j \xi^j) \partial_w + (\Xi_i^v-v_{j i}\xi^j )\partial^i_v  +(\Xi_i^w-w_{ji }\xi^j )\partial^i_w)  \rfloor  \\ \nonumber
& & ( [w^2  + \frac{1}{2}v_x^2+ aw_tv_{xx}-\frac{1}{2}av_{tt}v_{xx} -
 \frac{1}{2}bv_{xx}^2+\frac{1}{3}v_x^3+\frac{1}{2}\beta v_y^2]ds + ( \frac{1}{2}av_{xxt} 
 ) \omega^v\wedge ds_t  + \\ \nonumber
& & ( v_x+v_x^2-aw_{tx}+\frac{1}{2}av_{ttx}+bv_{xxx} ) \omega^v\wedge ds_x +  ( \beta v_y ) \omega^v\wedge ds_y + ( av_{xx} ) \omega^w\wedge ds_t + 
 \\ \nonumber
& &  ( -\frac{1}{2}av_{xx} ) \omega^v_t\wedge ds_t+ ( aw_t-\frac{1}{2}av_{tt}-bv_{xx} ) \omega^v_x\wedge ds_x +  \frac{1}{2}a\omega^w\wedge\omega^v_x\wedge ds_{tx})] 
\,.
\eEq
Now, comparing \eqref{first expression} with \eqref{straboh} we see that 
\bEq
d_H (J^{1}\Xi_V \rfloor p_{d_{V }\lambda} + \Xi_ H \rfloor \lambda) = d_H ( J^{1}\Xi\rfloor \rho_3)
\eEq
which implies 
\bEq 
& & d_H [(\xi^i(\partial_i+v_i \partial_v+ w_i \partial_w    + v_{ij}\partial_v^j + w_{ij } \partial_w^j )+
( (\Xi^v-v_j \xi^j) \partial_v +  (\Xi^w-w_j \xi^j) \partial_w + \nonumber
\\ 
& &  (\Xi_i^v-v_{j i}\xi^j)\partial^i_v  +
(\Xi_i^w-w_{ji}\xi^j )\partial^i_w)   \rfloor   \frac{1}{2}a\omega^w\wedge\omega^v_x\wedge ds_{tx}]  = 
 0 \,.
\eEq
By the exactness of the variational sequence, the above is equivalent to require that 
\bEq
 \frac{1}{2}a\,\,  (\xi^i \,  d_i + (\Xi^w-w_j \xi^j) \partial_w+ 
 (\Xi_x^v-v_{j x }\xi^j )\partial^x_v )  \rfloor  \omega^w\wedge\omega^v_x\wedge ds_{tx} = d_H \mu \,,
\eEq
\ie
\bEq
  \frac{1}{2}a\xi^y\omega^w\wedge\omega^v_x +   \frac{1}{2}a [ (\Xi^w-w_j \xi^j) \omega^v_x -(\Xi_x^v-v_{j x}\xi^j)\omega^w] \wedge ds_{tx} = d_H \mu \,.
\eEq
This equation specializes in our case of study the fact that, in general, there can be symmetry transformations of extremals which are not symmetries of equations, and this 
corresponds to the addition of a horizontal differential to the Noether--Bessel-Hagen current. 
\eEx

\section*{Acknowledgements} 

Research partially supported by Department of Mathematics - University of Torino through the projects ``Strutture geometriche e algebriche in fisica matematica e applicazioni" PALM$\_$RILO$\_20\_ 01$ and PALM$\_$RILO$\_22\_ 01$. 
This article is based upon work from COST Action 21109 CaLISTA, supported by
COST (European Cooperation in Science and Technology, {\em www.cost.eu}), partially {\em via} a Short-Term Scientific Mission (STSM) at the University of Brno, grant $e9840f91$.


\end{document}